\documentclass[journal]{IEEEtran}

\usepackage{tabularx}

\usepackage{booktabs}
\usepackage{multirow}
\newcommand{\head}[1]{\textnormal{\textbf{#1}}}

\usepackage{cite}                               %CITATION PACKAGES
\usepackage[cmex10]{amsmath}                    %MATH PACKAGES
\usepackage{amsthm}						  %Proof
\usepackage[pdftex]{graphicx}
\usepackage{url}                                %PDF, URL AND HYPERLINK PACKAGES ***
\usepackage{graphicx}                           %GRAPHICS RELATED PACKAGES ***
\usepackage{subfigure}                          %GRAPHICS RELATED PACKAGES ***
\usepackage{epstopdf}
\usepackage{amssymb}                            %For $\mathbf{R}$,$\mathbb{R}$
\usepackage{mdwmath}                            %Mark Wooding's tools for equation
\usepackage{threeparttable}                     %Table footnote
\usepackage{algorithmic}                        %ALGORITHMS
\usepackage[usenames, dvipsnames]{color}
\usepackage{empheq}
\usepackage{cancel}
\usepackage{soul}
\usepackage{colortbl}
\usepackage{array, indentfirst}
\ifCLASSINFOpdf
\else
\fi
\usepackage{tikz}

\usepackage{makecell}

\newcolumntype{?}{!{\vrule width 1.5pt}}
\usepackage{mathtools}

\DeclarePairedDelimiterX{\infdivx}[2]{(}{)}{%
	#1\;\delimsize\|\;#2%
}

\usepackage{scalerel,stackengine}
\stackMath
\newcommand\reallywidehat[1]{%
	\savestack{\tmpbox}{\stretchto{%
			\scaleto{%
				\scalerel*[\widthof{\ensuremath{#1}}]{\kern-.6pt\bigwedge\kern-.6pt}%
				{\rule[-\textheight/2]{1ex}{\textheight}}%WIDTH-LIMITED BIG WEDGE
			}{\textheight}% 
		}{0.5ex}}%
	\stackon[1pt]{#1}{\tmpbox}%
}

\usepackage{enumerate}
\usepackage{hhline}
\newcommand{\rom}[1]{\uppercase\expandafter{\romannumeral#1\relax}}

\usepackage{color, colortbl} 
\definecolor{Gray}{gray}{0.9}
\usepackage{hyperref}

\newcommand{\norm}[1]{\left\lVert#1\right\rVert}

\begin{document}
	\title{Low computational cost method for online parameter identification of Li-ion battery in battery management systems using matrix condition number  }
	
	\author{
		\vskip 1em
		
		%		First A. Author1, \emph{Student Membership},
		%		Second B. Author2, \emph{Membership},
		%		\\ and Third C. Author3, \emph{Membership}
		Minho Kim, Kwangrae Kim, Soohee Han*
		
		\thanks{

			(Authors' names and affiliation) Minho Kim and Soohee Han are with the Creative IT Engineering Department, Pohang University of Science and Technology, Pohang, Republic of Korea.
			(email: [minho1st, soohee.han]@postech.ac.kr). Kwangrae Kim is with the Electrical Engineering Department, Pohang University of Science and Technology, Pohang, Republic of Korea. (email: caritas@postech.ac.kr).

		}
	}
	
	%	\author{Minho Kim, Soohee Han*}  
	
	\maketitle
	
	\begin{abstract}
	Monitoring the state of health for Li-ion batteries is crucial in the battery management system (BMS), which helps end-users use batteries efficiently and safely. Battery state of health can be monitored by identifying parameters of battery models using various algorithms. Due to the low computation power of BMS and time-varying parameters, it is very important to develop an online algorithm with low computational cost. Among various methods, Equivalent circuit model (ECM) -based recursive least squares (RLS) parameter identification is well suited for such difficult BMS environments. However, one well-known critical problem of RLS is that it is very likely to be numerically unstable unless the measured inputs make enough excitation of the battery models. In this work, A new version of RLS, which is called condition memory recursive least squares (CMRLS) is developed for the Li-ion battery parameter identification to solve such problems and to take advantage of RLS at the same time by varying forgetting factor according to condition numbers. In CMRLS, exact condition numbers are monitored with simple computations using recursive relations between RLS variables. The performance of CMRLS is compared with the original RLS through Li-ion battery simulations. It is shown that CMRLS identifies Li-ion battery parameters about 100 times accurately than RLS in terms of mean absolute error.

	\end{abstract}
	
	\begin{IEEEkeywords}
		Battery management system, Condition memory recursive least squares, Condition number, Recursive least squares.
	\end{IEEEkeywords}
	
%	\markboth{IEEE TRANSACTIONS ON INDUSTRIAL INFORMATICS}%
	{}
	
	\definecolor{limegreen}{rgb}{0.2, 0.8, 0.2}
	\definecolor{forestgreen}{rgb}{0.13, 0.55, 0.13}
	\definecolor{greenhtml}{rgb}{0.0, 0.5, 0.0}
	
	\section{Introduction}
	Li-ion batteries have been used in many applications. Accordingly, for efficient and safe battery usage, monitoring state of health (SOH) and state of charge (SOC) has been very important in the battery management system (BMS). In the industrial and academic fields, battery parameters such as capacity and internal resistance have been used as SOH very commonly, which is very crucial parameters for end-users. These parameter values can also used to estimate SOC because the parameters determine the dynamics of SOC. Therefore, monitoring battery parameters is crucial for SOH and SOC estimation and thus for BMS. This can be done by applying parameter estimation algorithms to mathematical models for Li-ion batteries. 
	
	Li-ion battery parameter identification methods can be categorized into two classes, methods based on electrochemical battery models and those based on equivalent circuit models (ECMs). 
		
   	Many researchers have studied parameter identification methods for the electrochemical battery models. Electrochemical battery model parameter identifications have been mostly carried out using meta-heuristic algorithms such as genetic algorithms\cite{forman2011genetic,li2016parameter,forman2012genetic}, adaptive exploration harmony search\cite{chun2019adaptive} ,and particle swarm optimization\cite{rahman2016electrochemical} because meta-heuristic algorithm have a capability to find a global optima of such complex battery models consisting of nonlinear algebraic equations. The meta-heuristic algorithms, however, take a lot of time because of their repeated error evaluation process until the convergence. To solve this problem, some researchers have used machine learning techniques for the parameter identification of electrochemical models\cite{kim2019data,chun2019parameter,kimj2019data}. However, trained machine learning models is too heavy to be embedded on the BMS microprocessor chips that is usually made cheap for competitiveness in the markets related to the Li-ion batteries.
   	For real-time parameter monitoring for the Li-ion batteries in the BMSs, light and fast parameter identification algorithms are need.
   	
   	Parameter identification based on ECMs are more suitable than electrochemical models for such purposes because parameter identification algorithms for electrochemical models are very likely to be heavy or time-consuming to deal with complex nonlinear algebraic equations of them, which is not the case for ECMs consisting of simple linear equations. ECM-based parameter identification has been usually carried out using Kalman filter (KF)-based algorithms because most ECMs consist of linear equations \cite{hu2012multiscale,plett2004extended,dai2009state,do2009impedance}. KF-based methods are specialized to track parameters varying fast due to their mathematically well organized prediction and update phases of each iteration. Although KF-based methods are more simple and practical than above-mentioned electrochemical model-based methods from the perspective of usage in the BMSs, there are still heavy computations such as matrix inversion and tuning of covariance matrices. Recursive least squares (RLS) is a better solution because it does not need to inverse matrices or to tune covariance matrices. Although KF-based methods can track fast-varying parameters better than RLS, RLS can show similar accuracy compared to KF-based methods in the battery parameter identification because the dynamics of the battery parameters are usually slow. 
   	
   	For these reasons, RLS-based algorithms have been used in the battery parameter identifications\cite{hu2011online,duong2015online,he2012online,zhang2018online}. However, one big problem of RLS-based algorithms is their well-known numerical instability problem, which is called a wind-up problem. The wind-up problem is caused when excitation of the system dealt with is poor and thus the covariance matrix of the parameter estimation in the RLS become very large\cite{fortescue1981implementation}, which leads to very high sensitivity of solved parameter values to the truncation error and the error of sensors. To solve the wind-up problem, some researchers have developed modified version of RLS by using a variable forgetting factor according to the parameter estimation errors\cite{fortescue1981implementation,leung2005gradient} or to the trace of the covariance matrix\cite{rao1987improved}. However, these methods do not measure numerical stability directly. One of the well-known direct measurement of numerical stability of the parameter identification is measure condition number of the covariance matrix.
   	
   	In this paper, a new version of RLS, called condition memory recursive least squares (CMRLS), is developed for real-time Li-ion battery equivalent circuit model (ECM) parameter identification in the BMS hardware with low computing power. The newly developed CMRLS adaptively changes a forgetting factor according to the condition number of the covariance matrix so that the wind-up problem is prevented. Computing a condition number of a matrix requires an inversion operation but CMRLS obtains the condition number with a simpler computation using recursive relations between the variables of RLS. The validation of CMRLS is carried out with a ECM simulation data. The proposed CMRLS shows about 100 times more accurate parameter identification than the original RLS in terms of mean absolute error.

    \section{Matrix condition number}\label{condition_number}
    
	The matrix conditon number of a square matrix $ A\in \mathbb{R}^{n\times n} $  is defined as follows:
	\begin{eqnarray}
	 \kappa(A) = \norm{A} \norm{A^{-1}} 
	\end{eqnarray}
	\noindent where $ \norm{A} $ means the matrix norm of $ A $. If $ x $ is a solution of a linear equation
	\begin{eqnarray}
	Ax=b
	\end{eqnarray}
	where $ b\in \mathbb{R}^{n\times 1} $ and $ x+\Delta x $ is a solution of a linear equation
	\begin{eqnarray}
	(A+\Delta A)(x+\Delta x)=(b+\Delta b),
	\end{eqnarray}
	it can be shown that the following inequality holds with some assumptions:
	\begin{eqnarray}
	\frac{\frac{\norm{\Delta x}}{\norm{x}}}{\frac{\norm{\Delta b}}{\norm{b}}+\frac{\norm{\Delta A}}{\norm{A}}} \leq \kappa(A)
	\label{cond_inequality}
	\end{eqnarray}
	Therefore, $ \kappa(A) $ is a measurement of how sensitively the solution of a linear equation $ Ax=b $ changes according to the perturbation of $ A $ and $ b\in \mathbb{R}^{n\times 1} $. The proof of (\ref{cond_inequality}) is presented in \cite{conditon_proof}.

	\section{Recursive least squares}
	
	Recursive least squares (RLS) is an recursive algorithm for solving the least squares (LS) problem of finding the parameters $ \theta \in \mathbb{R}^{n\times 1} $ of a linear regression model
	\begin{eqnarray}
	d_t=\theta^T\phi_t
	\label{linear}
	\end{eqnarray}
	
	where  $ d \in \mathbb{R}^{1\times 1} $ is an output and $ \phi \in \mathbb{R}^{n\times 1} $ is an input of the linear model when time series data $ [\phi_0 , d_0],[\phi_1 , d_1],[\phi_2 , d_2],...,[\phi_t , d_t],... $  are given. Specifically, the problem solved by RLS is to find $ \theta $ minimizing the output error $ e(t) $ at time step $ t $ defined as
	\begin{eqnarray}
	e(t)=\sum_{k=0}^{t}\lambda^{t-k}(d_k-\theta^T\phi_k)^2
	\label{error}
	\end{eqnarray}
	where $ \lambda $ is a real number between 0 and 1, which is called the forgetting factor. $ \lambda $ assigns bigger weights to the present data compared to the past data when computing the output error. With small lambda, a large amount of the past data is forgotten when finding optimal $ \theta $ so that RLS can deal with time-varying $ \theta $. It can be easily shown that the optimal solution $ \theta_{t} $ minimizing $ e(t) $ is a solution of the following linear equation:
	\begin{eqnarray}
	\Phi_{t}\theta_{t}=\Psi_{t}
	\label{linear_eq}
	\end{eqnarray}
	where 
	\begin{eqnarray}
	\Phi_{t}=\sum_{k=0}^{t}\lambda^{t-k}\phi_k\phi_k^T\\
	\Psi_{t}=\sum_{k=0}^{t}\lambda^{t-k}\phi_kd_k
	\end{eqnarray}
	Applying the matrix inversion formula, the LS problem (\ref{linear_eq}) can be formulated as a recursive algorithm which is called RLS expressed as following :
	\begin{eqnarray}
	k_{t}&\leftarrow&\frac{P_{t-1}\phi_t}{\lambda+\phi_{t}^TP_{t-1}\phi_t}\\
	\alpha_{t}&\leftarrow& d_t-\phi_t^TP_{t-1}\\
	P_{t}&\leftarrow& \frac{P_{t-1}-k_t\phi_t^TP_{t-1}}{\lambda}\\
	\theta_t &\leftarrow& \theta_{t-1} + k_t\alpha_t
	\end{eqnarray}
	where $ P_t =[\Phi_t]^{-1}$ is covariance matrix for parameter $ \theta $ estimation from the perspective of Kalman filter \cite{soderstrom1988system}.

	\section{Condition memory recursive least squares for the parameter identification of the Li-ion batteries}
	
	\subsection{Equivalent circuit model of the Li-ion batteries}
	
	\begin{figure}[htbp]\centering
		\includegraphics[width=0.5\textwidth]{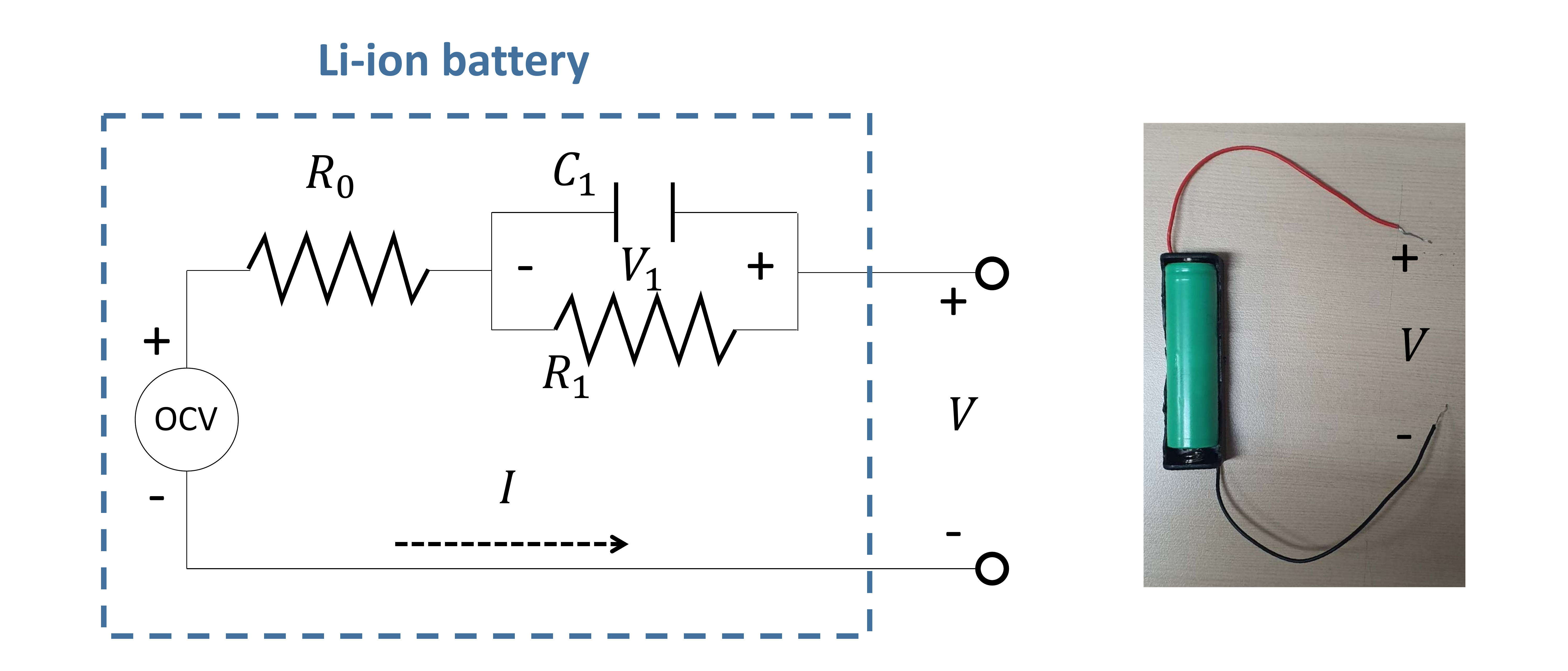}
		\caption{1 RC equivalent circuit model (ECM)}\label{1RC_circuit}
	\end{figure}
\begin{figure}[htbp]\centering
		\includegraphics[width=0.5\textwidth]{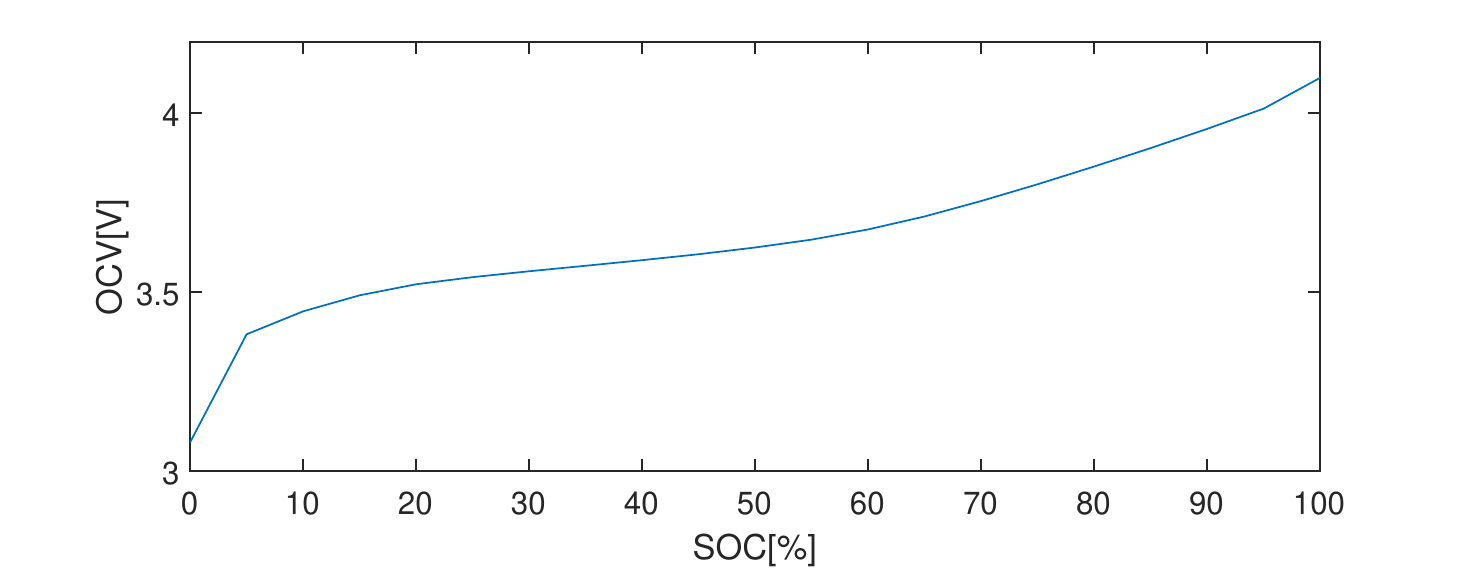}
		\caption{SOC-OCV relationship obtained with the experiments.}\label{SOCOCVtable}
	\end{figure}

	In this paper, One of the popular equivalent circuit models (ECMs) called ``1RC model'' is used, which is shown in Figure \ref{1RC_circuit} where $ I $ is current and $ V $ is terminal voltage of the battery. OCV means open circuit voltage which is assumed to be a function of SOC. The OCV-SOC relationship used in this paper is obtained from slow charge or discharge experiments(Figure \ref{SOCOCVtable}).
	OCV is assumed to be linear function of SOC, which means $ \text{OCV} = \beta_1\text{SOC }+\beta_2$ where $ \beta_1 $ and $ \beta_2 $ are parameters of the ECM. Although this assumption seems unrealistic, it does not prevent the newly developed CMRLS from identifying accurate parameter values because the RLS-based methods can deal with time-varying parameters and $ \beta_1 $ and $ \beta_2 $ vary with time in the real world. Setting $ x=\left[ {\begin{array}{c}
		V_1\\
		\text{SOC} \\
		\end{array} } \right] $ as a state vector, a state-space model for this ECM can be expressed as follows:
	\begin{eqnarray}
	\frac{d}{dt}x=Ax+Bu \label{continuous_state_space_model_1}\\ 
	y=Cx+Du \label{continuous_state_space_model_2}
	\end{eqnarray}
	where 
	\begin{eqnarray*}
	A=\left[ {\begin{array}{cc}
		-\frac{1}{R_1C_1}&0\\
		0&0 \\
		\end{array} } \right], 
	B=\left[ {\begin{array}{c}
		\frac{1}{C_1}\\
		\frac{1}{Cap} \\
		\end{array} } \right],	
	C=\left[ {\begin{array}{cc}
			1&\beta_1\\
	\end{array} } \right],
	\end{eqnarray*}

	\begin{eqnarray*}
	D=R_0, y=V-\beta_2, u=I
	\end{eqnarray*}
	
	$ Cap $ is the capacity of the battery. The state-space model in (\ref{continuous_state_space_model_1}) and (\ref{continuous_state_space_model_2}) can be converted into discrete-time state-space model as follows:
	
	\begin{eqnarray}
	x_{t+1}=A_dx_{t}+B_du_t \label{discrete_state_space_model_1}\\ 
	y_t=C_dx_t+D_du_t \label{discrete_state_space_model_2}
	\end{eqnarray}
	where 
	\begin{eqnarray*}
		A_d&=&e^{\Delta tA}=\left[ {\begin{array}{cc}
				\exp{\left(-\frac{\Delta t}{R_1C_1}\right)}&0\\
				0&1 \\
		\end{array} } \right],\\
		B_d&=&\left(\int_{0}^{\Delta t}e^{At}dt\right)B=\left[ {\begin{array}{c}
				R_1\left(1-\exp{\left(-\frac{\Delta t}{R_1C_1}\right)}\right)\\
				\frac{\Delta t}{Cap} \\
		\end{array} } \right],\\
		C_d&=&C=\left[ {\begin{array}{cc}
				1&\beta_1\\
		\end{array} } \right],\\
		D_d&=&D=R_0,\\
		y_t&=&V_t-\beta_2, u_t=I_t
	\end{eqnarray*}
	where $ \Delta t $ is a time step.
	Applying z-transform to the discrete-time state-space model equations in (\ref{discrete_state_space_model_1}) and (\ref{discrete_state_space_model_2}) can be represented as follows:
	\begin{eqnarray}
	\frac{V(z)-\beta_2}{I(z)}&=&C_d(zI-A_d)^{-1}B_d+D_d \nonumber\\
	&=&\frac{a_3+a_4z^{-1}+a_5z^{-2}}{1+a_1z^{-1}+a_2z^{-2}} \label{z_transform}
	\end{eqnarray}
	where $ V(z) $ is the z-transform of $ V_t $, $ I(z) $ is the z-transform of $ I_t $ and
	\begin{eqnarray*}
	a_1&=&-\exp{\left(-\frac{\Delta t}{R_1C_1}\right)}-1,\\
	a_2&=&\exp{\left(-\frac{\Delta t}{R_1C_1}\right)},\\
	a_3&=&R_0,\\
	a_4&=&R_1-R_1a_2+\frac{\beta_1\Delta t}{Cap}-R_0a_2-R_0,\\
	a_5&=&-R_1+R_1a_2-\frac{a_2\beta_1\Delta t}{Cap}+R_0a_2
	\end{eqnarray*}
	Using (\ref{z_transform}), a linear regression model can be formulated as follows:
	\begin{eqnarray}
	V_t-V_{t-1}=\left[ {\begin{array}{c}
		a_2\\
		a_3\\
		a_4\\
		a_5\\
		\end{array} } \right]^T
	\left[ {\begin{array}{c}
		V_{t-1}-V_{t-2}\\
		I_t\\
		I_{t-1}\\
		I_{t-2}\\
		\end{array} } \right]
	\end{eqnarray}
	where
	$ V_t-V_{t-1} $, $ \left[ {\begin{array}{c}
		a_2\\
		a_3\\
		a_4\\
		a_5\\
		\end{array} } \right] $, and $ \left[ {\begin{array}{c}
		V_{t-1}-V_{t-2}\\
		I_t\\
		I_{t-1}\\
		I_{t-2}\\
		\end{array} } \right] $ corresponds to $ d_t $, $ \theta $, and $ \phi_t $ in (\ref{linear}), respectively.
	
	\subsection{Parameter identification using condition memory recursive least squares}
\begin{figure}[htbp]\centering
		\includegraphics[width=0.31\textwidth]{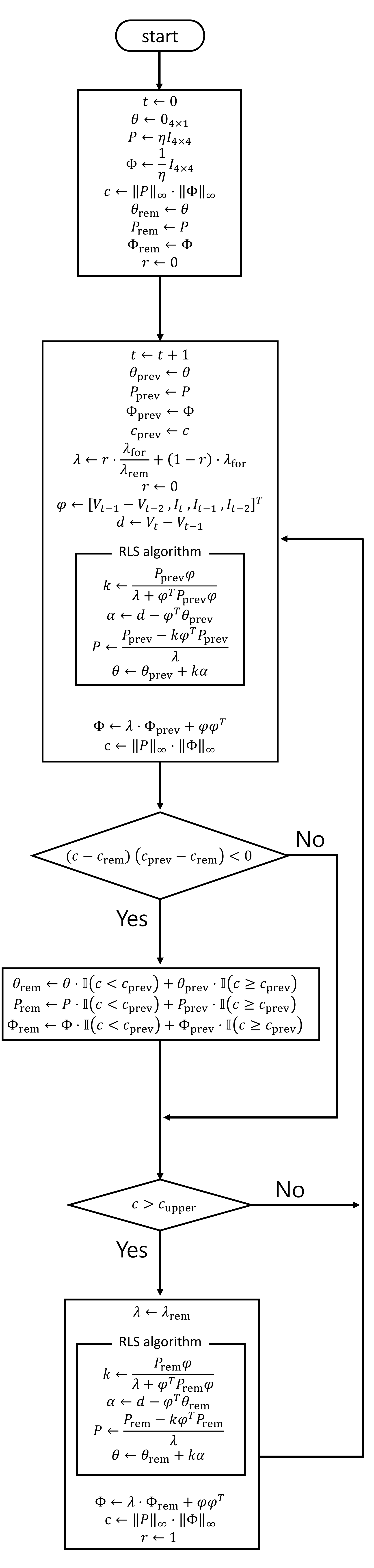}
		\caption{Flow chart of the proposed CMRLS. $ c $ is condition number of the covariance matrix $ P $ or $ \kappa(P) $}\label{CMRLS_flow_chart}
	\end{figure}

	 The strategy of the newly proposed parameter identification algorithm, which is called condition memory recursive least squares(CMRLS) in this paper, is simple.
	 As mentioned in the section \ref{condition_number}, when given a linear equation $ Ax=b $, small condition number of $ A $ means high numerical stability for the solution of a linear equation. Because the RLS solves the linear equation in (\ref{linear_eq}), the condition number of covariance matrix $ P_t $ is the important number determining how much numerically stable the solution $ theta_t $ is.   Therefore, CMRLS memorizes RLS variables ($ k_t, \alpha_t, P_t, \theta_t $) when $ \kappa(P_t) $ is small, which are used when  $ \kappa(P_t) $ becomes too big so that CMRLS continuously outputs numerically stable and reliable solutions of $ \theta_t $. The flow chart for CMRLS is shown in Figure \ref{CMRLS_flow_chart}.
	
	There are several design parameters of the CMRLS:
	\begin{itemize}
		\item $ c_{\text{rem}} $: RLS variables are memorized $ \kappa(P) $ crosses this number. this number have to be set small enough for the CMRLS memorize to variables which are numerically stable enough.
		\item $ c_{\text{upper}} $: Memorized RLS variables is used to obtain $ \theta_t $ when  $ \kappa(P_t) $ becomes bigger than this number. If this number is small, the solution $ \theta_t $ is very likely to be stable because memorized RLS variables are frquently used but parameter tracking performance is bad because the past data is frequently used. Therefore, this number have to be tuned properly considering this trade-off.
		\item $ \lambda_{\text{for}} $: This number is the same as forgetting factor in the original RLS.
		\item $ \lambda_{\text{rem}} $: This number is set to be bigger than 1. The forgetting factor is replaced with this number to give large weight to the memorized RLS variables when they are used to obtain the solution $ \theta_t $. 
	\end{itemize}
	
	Infinity norm is ued to calculate the condition number. One of the remarkable things of CMRLS is that condition number computation does not require matrix inversion which have high computational complexity. Although the condition number computation originally requires matrix inversion, in CMRLS, using the relation between $ P $ and $ \Phi $ ($ P^{-1}=\Phi $), the computation of the condition number $ \kappa(P) $ does not require matrix inversion as follows:
	\begin{eqnarray}
	\kappa(P)=\norm{P}\norm{P^{-1}}=\norm{P}\norm{\Phi}
	\end{eqnarray}
	Therefore, CMRLS can be practically used in the BMS with low computation power hardware.

	\section{Results and discussions}
\begin{figure}[htbp]\centering
		\includegraphics[width=0.35\textwidth]{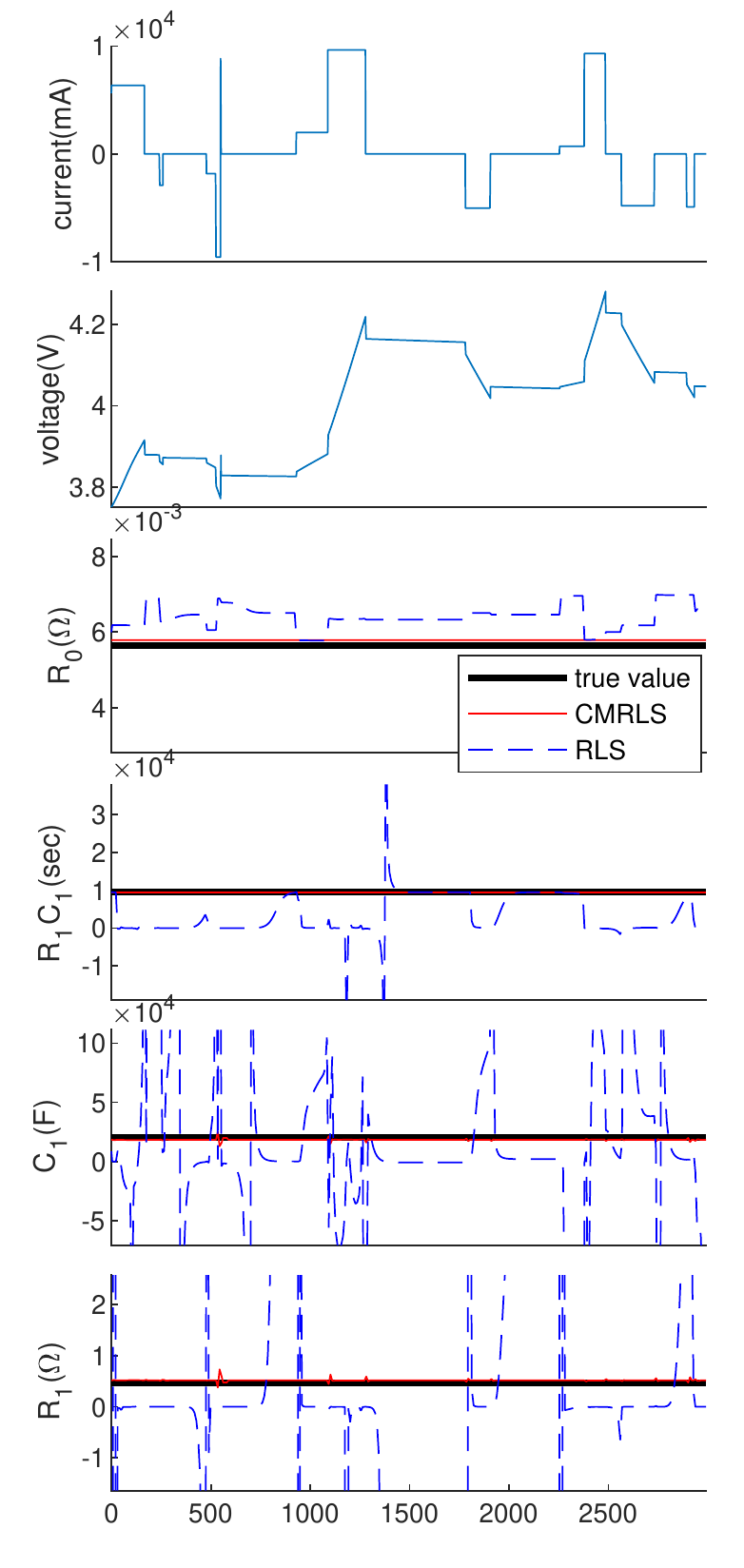}
		\caption{Valdiation result of the proposed CMRLS.}\label{val_1}
	\end{figure}

\begin{figure}[htbp]\centering
		\includegraphics[width=0.35\textwidth]{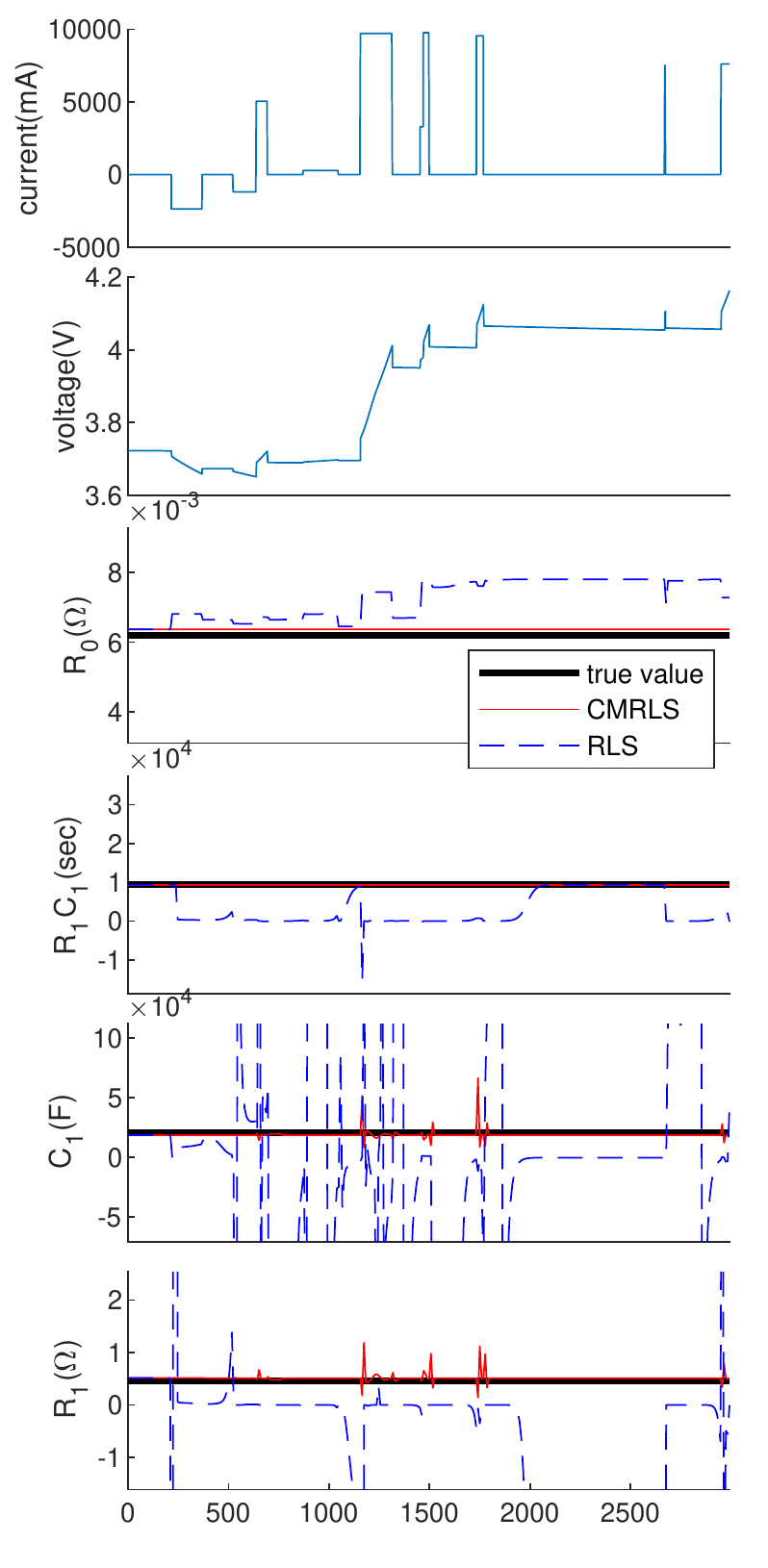}
		\caption{Valdiation result of the proposed CMRLS.}\label{val_2}
	\end{figure}

	Not only can the proposed CMRLS be practically used in the BMS, but it is shown that it also has high accuracy in identifying parameters. The validation is carried out using random pulse simulation data(Figure \ref{val_1} and Figure \ref{val_2}). CMRLS is carried out with time step of 10 sec which is 1000 times longer than the simulation model computation time step. Considering such long time step for the CMRLS time step, it can be known that the proposed CMRLS can identify battery parameters accurately even with low sampling frequency, which motivates to use CMRLS in the BMS.
	

	\begin{table}[h!]
		\centering
		\begin{tabularx}{0.48\textwidth}{XXXXXX}%
			\toprule[1.5pt]		
			& & \multicolumn{2}{c}{\head{RLS}}
			& \multicolumn{2}{c}{\head{CMRLS}}\\
			\cmidrule(lr){3-4}\cmidrule(l){5-6}\\
			
			& true value & mean identified value & mean abs error &
			mean identified value & mean abs error\\
			\cmidrule(lr){2-2}\cmidrule(lr){3-4}\cmidrule(l){5-6}\\
			$ R_0 \ \ [\Omega] $ & $ 6.193\times 10^{-3} $ & $ 7.243\times 10^{-3} $ & $ 1.049\times 10^{-3} $ & $ 6.368\times 10^{-3} $ & $ 1.749\times 10^{-4} $ \\
			\cmidrule(lr){2-2}\cmidrule(lr){3-4}\cmidrule(l){5-6}\\
			$ R_1C_1 [sec] $ & $ 9.359\times 10^{3} $ & $ 3.194\times 10^{3} $ & $ 6.165\times 10^{3} $ & $ 9.459\times 10^{3} $ & $ 5.928\times 10^{-5} $ \\
			\cmidrule(lr){2-2}\cmidrule(lr){3-4}\cmidrule(l){5-6}\\
			$ C_1 \ \ [F]$ & $ 2.029\times 10^{4} $ & $ 1.7484\times 10^{4} $ & $ 2.396\times 10^{5} $ & $ 1.854\times 10^{4} $ & $ 2.375\times 10^{3} $ \\
			\cmidrule(lr){2-2}\cmidrule(lr){3-4}\cmidrule(l){5-6}\\
			$ R_1 \ \ [\Omega]$ & $ 4.613\times 10^{-1} $ & $ -5.741\times 10^{0} $ & $ 1.085\times 10^{1} $ & $ 5.146\times 10^{-1} $ & $ 5.972\times 10^{-2} $ \\
			\bottomrule[1.5pt]\\
		\end{tabularx}
		
		\caption{Valdiation results of the proposed CMRLS}
		\label{validation_result_table}
	\end{table}

	The performance of the proposed CMRLS is compared with original RLS algorithm. As shown in Figure \ref{val_1} and Figure \ref{val_2}, CMRLS has much less parameter identification error than RLS.
	Table \ref{validation_result_table} represents mean estimated values and mean absolute errors of RLS and CMRLS for the result show in Figure \ref{val_2}.

	\section{Conclusion}
	In this paper, a new version of RLS, which is called condition memory recursive least squares (CMRLS), is proposed to identify the parameters of Li-ion batteries. The proposed CMRLS can accurately identify parameters with low computational complexity compared to the other parameter identification algorithms, which is very suitable for the usage in the commercial BMS hardware with low computing power. CMRLS solves a numerical stability problem of the original RLS, one of the critical problems of it. One of the limitation is that this algorithm cannot reflect the real physical phenomena inside the Li-ion batteries. Therefore, the future work will be to identify parameters which have much more physical meanings then ECM parameters using data-driven algorithms.

%	\section{Appendix}
%	
%	\setcounter{equation}{0}
%	\renewcommand{\theequation}{A.\arabic{equation}}
%	
%	\subsection{Proof of } \label{Appendix_1}
%	dsafsdaf
%	\begin{eqnarray}
%	\norm{A}
%	\end{eqnarray}

    \bibliographystyle{Bibliography/IEEEtran}
    
    \bibliography{Bibliography/IEEEabrv,Bibliography/MINHOBIB}

\begin{thebibliography}{10}
\providecommand{\url}[1]{#1}
\csname url@samestyle\endcsname
\providecommand{\newblock}{\relax}
\providecommand{\bibinfo}[2]{#2}
\providecommand{\BIBentrySTDinterwordspacing}{\spaceskip=0pt\relax}
\providecommand{\BIBentryALTinterwordstretchfactor}{4}
\providecommand{\BIBentryALTinterwordspacing}{\spaceskip=\fontdimen2\font plus
\BIBentryALTinterwordstretchfactor\fontdimen3\font minus
  \fontdimen4\font\relax}
\providecommand{\BIBforeignlanguage}[2]{{%
\expandafter\ifx\csname l@#1\endcsname\relax
\typeout{** WARNING: IEEEtran.bst: No hyphenation pattern has been}%
\typeout{** loaded for the language `#1'. Using the pattern for}%
\typeout{** the default language instead.}%
\else
\language=\csname l@#1\endcsname
\fi
#2}}
\providecommand{\BIBdecl}{\relax}
\BIBdecl

\bibitem{forman2011genetic}
J.~C. Forman, S.~J. Moura, J.~L. Stein, and H.~K. Fathy, ``Genetic parameter
  identification of the {D}oyle-{F}uller-{N}ewman model from experimental
  cycling of a {L}i{F}e{PO}$_4$ battery,'' in \emph{American Control Conference
  (ACC), 2011}.\hskip 1em plus 0.5em minus 0.4em\relax IEEE, 2011, pp.
  362--369.

\bibitem{li2016parameter}
J.~Li, L.~Zou, F.~Tian, X.~Dong, Z.~Zou, and H.~Yang, ``Parameter
  identification of lithium-ion batteries model to predict discharge behaviors
  using heuristic algorithm,'' \emph{Journal of The Electrochemical Society},
  vol. 163, no.~8, pp. A1646--A1652, 2016.

\bibitem{forman2012genetic}
J.~C. Forman, S.~J. Moura, J.~L. Stein, and H.~K. Fathy, ``Genetic
  identification and fisher identifiability analysis of the
  {D}oyle--{F}uller--{N}ewman model from experimental cycling of a
  {L}i{F}e{PO}$_4$ cell,'' \emph{Journal of Power Sources}, vol. 210, pp.
  263--275, 2012.

\bibitem{chun2019adaptive}
H.~Chun, M.~Kim, J.~Kim, K.~Kim, J.~Yu, T.~Kim, and S.~Han, ``Adaptive
  exploration harmony search for effective parameter estimation in an
  electrochemical lithium-ion battery model,'' \emph{IEEE Access}, vol.~7, pp.
  131\,501--131\,511, 2019.

\bibitem{rahman2016electrochemical}
M.~A. Rahman, S.~Anwar, and A.~Izadian, ``Electrochemical model parameter
  identification of a lithium-ion battery using particle swarm optimization
  method,'' \emph{Journal of Power Sources}, vol. 307, pp. 86--97, 2016.

\bibitem{kim2019data}
M.~Kim, H.~Chun, J.~Kim, K.~Kim, J.~Yu, T.~Kim, and S.~Han, ``Data-efficient
  parameter identification of electrochemical lithium-ion battery model using
  deep bayesian harmony search,'' \emph{Applied Energy}, vol. 254, p. 113644,
  2019.

\bibitem{chun2019parameter}
H.~Chun, J.~Kim, and S.~Han, ``Parameter identification of an electrochemical
  lithium-ion battery model with convolutional neural network,''
  \emph{IFAC-PapersOnLine}, vol.~52, no.~4, pp. 129--134, 2019.

\bibitem{kimj2019data}
J.~Kim, H.~Chun, M.~Kim, J.~Yu, K.~Kim, T.~Kim, and S.~Han, ``Data-driven state
  of health estimation of li-ion batteries with rpt-reduced experimental
  data,'' \emph{IEEE Access}, vol.~7, pp. 106\,987--106\,997, 2019.

\bibitem{hu2012multiscale}
C.~Hu, B.~D. Youn, and J.~Chung, ``A multiscale framework with extended kalman
  filter for lithium-ion battery soc and capacity estimation,'' \emph{Applied
  Energy}, vol.~92, pp. 694--704, 2012.

\bibitem{plett2004extended}
G.~L. Plett, ``Extended kalman filtering for battery management systems of
  lipb-based hev battery packs: Part 3. state and parameter estimation,''
  \emph{Journal of Power sources}, vol. 134, no.~2, pp. 277--292, 2004.

\bibitem{dai2009state}
H.~Dai, X.~Wei, and Z.~Sun, ``State and parameter estimation of a hev li-ion
  battery pack using adaptive kalman filter with a new soc-ocv concept,'' in
  \emph{2009 International Conference on Measuring Technology and Mechatronics
  Automation}, vol.~2.\hskip 1em plus 0.5em minus 0.4em\relax IEEE, 2009, pp.
  375--380.

\bibitem{do2009impedance}
D.~V. Do, C.~Forgez, K.~E.~K. Benkara, and G.~Friedrich, ``Impedance observer
  for a li-ion battery using kalman filter,'' \emph{IEEE Transactions on
  Vehicular Technology}, vol.~58, no.~8, pp. 3930--3937, 2009.

\bibitem{hu2011online}
X.~Hu, F.~Sun, Y.~Zou, and H.~Peng, ``Online estimation of an electric vehicle
  lithium-ion battery using recursive least squares with forgetting,'' in
  \emph{Proceedings of the 2011 American Control Conference}.\hskip 1em plus
  0.5em minus 0.4em\relax IEEE, 2011, pp. 935--940.

\bibitem{duong2015online}
V.-H. Duong, H.~A. Bastawrous, K.~Lim, K.~W. See, P.~Zhang, and S.~X. Dou,
  ``Online state of charge and model parameters estimation of the lifepo4
  battery in electric vehicles using multiple adaptive forgetting factors
  recursive least-squares,'' \emph{Journal of Power Sources}, vol. 296, pp.
  215--224, 2015.

\bibitem{he2012online}
H.~He, X.~Zhang, R.~Xiong, Y.~Xu, and H.~Guo, ``Online model-based estimation
  of state-of-charge and open-circuit voltage of lithium-ion batteries in
  electric vehicles,'' \emph{Energy}, vol.~39, no.~1, pp. 310--318, 2012.

\bibitem{zhang2018online}
C.~Zhang, W.~Allafi, Q.~Dinh, P.~Ascencio, and J.~Marco, ``Online estimation of
  battery equivalent circuit model parameters and state of charge using
  decoupled least squares technique,'' \emph{Energy}, vol. 142, pp. 678--688,
  2018.

\bibitem{fortescue1981implementation}
T.~Fortescue, L.~S. Kershenbaum, and B.~E. Ydstie, ``Implementation of
  self-tuning regulators with variable forgetting factors,'' \emph{Automatica},
  vol.~17, no.~6, pp. 831--835, 1981.

\bibitem{leung2005gradient}
S.-H. Leung and C.~So, ``Gradient-based variable forgetting factor rls
  algorithm in time-varying environments,'' \emph{IEEE Transactions on Signal
  Processing}, vol.~53, no.~8, pp. 3141--3150, 2005.

\bibitem{rao1987improved}
N.~Rao~Sripada and D.~Grant~Fisher, ``Improved least squares identification,''
  \emph{International Journal of Control}, vol.~46, no.~6, pp. 1889--1913,
  1987.

\bibitem{conditon_proof}
\BIBentryALTinterwordspacing
 [Online]. Available:
  \url{https://www2.math.ethz.ch/education/bachelor/lectures/hs2014/other/linalg_INFK/matrix-condition-number.pdf}
\BIBentrySTDinterwordspacing

\bibitem{soderstrom1988system}
T.~S{\"o}derstr{\"o}m and P.~Stoica, \emph{System identification}.\hskip 1em
  plus 0.5em minus 0.4em\relax Prentice-Hall, Inc., 1988.

\end{thebibliography}

\end{document}